\documentclass[10pt]{article}

\usepackage{amsmath}
\usepackage{array}
\usepackage{appendix}
\usepackage{graphicx}
\usepackage{amsfonts}
\usepackage{amssymb}
\usepackage{mathrsfs}
\usepackage{yfonts}
\usepackage{euscript}
\usepackage{upgreek}
\usepackage{slantsc}
\usepackage{calligra}
\usepackage[T1]{fontenc}
\usepackage{epsf}
\usepackage{latexsym}

\usepackage{tipa}

\def\be{\begin{equation}}
\def\ee{\end{equation}}
\def\beq{\begin{equation}}
\def\eeq{\end{equation}}
\def\bea{\begin{eqnarray}}
\def\eea{\end{eqnarray}}

\def\ni{\noindent}
\def\foo{\footnote}

\def\!{\hspace{-1.6667em}}

\def\mD{\mbox{D}}

\def\mN{\mbox{N}}

\def\mR{\mbox{R}}

\def\mV{\mbox{V}}

\def\md{\mbox{d}} 
\def\me{\mbox{e}}

\def\mh{\mbox{h}}

\def\mp{\mbox{p}}

\def\brho{\mbox{\boldmath$\rho$}}          % mass-weighted relative Jacobi configuration space vector
\def\fG{\mbox{\sffamily G}}

\def\fQ{\mbox{\sffamily Q}}
\def\fR{\mbox{\sffamily R}}
\def\fS{\mbox{\sffamily S}}

\def\bn{\mbox{\bf n}}

\def\bp{\mbox{\bf p}}
\def\bA{\mbox{\bf A}}
\def\bB{\mbox{\bf B}}

\def\bn{\mbox{{\bf n}}}
\def\bq{\mbox{{\bf q}}}

\def\bV{\mbox{\bf V}}

\def\FrMgen{\mbox{\boldmath$\mathfrak{M}$}}                              % The general manifold eg in the theory of foliations
\def\scc{\mbox{\scriptsize c}}

\def\se{\mbox{\scriptsize e}}

\def\sll{\mbox{\scriptsize l}}  %NB EXCEPTIONAL DEF as \sl is reserved for slant.
\def\sm{\mbox{\scriptsize m}}
\def\sn{\mbox{\scriptsize n}} 
\def\so{\mbox{\scriptsize o}} 
\def\sp{\mbox{\scriptsize p}}

\def\sss{\mbox{\scriptsize s}}  %TO AVOID ARXIV changing \ss to German double s.

\def\sA{\mbox{\scriptsize A}} 
\def\sB{\mbox{\scriptsize B}}

\def\sP{\mbox{\scriptsize P}} 
 
\def\sR{\mbox{\scriptsize R}}

\def\sbA{\mbox{{\bf \scriptsize A}}}
\def\sbB{\mbox{{\bf \scriptsize B}}}

\def\5Star{\mbox{\Large$\star$}}              % big five-point star for hanging things on
\def\cr{\mbox{\scriptsize{\bf $\mbox{ } \times \mbox{ }$}}}

\def\sumi3{\sum\mbox{}_{\mbox{}_{\mbox{\scriptsize $i$=1}}}^3}

\def\sumIN{\sum\mbox{}_{\mbox{}_{\mbox{\scriptsize $I$=1}}}^{N}}
\def\sumj3{\sum\mbox{}_{\mbox{}_{\mbox{\scriptsize $j$=1}}}^3}
\def\sumk3{\sum\mbox{}_{\mbox{}_{\mbox{\scriptsize $k$=1}}}^3}

\begin{document}

\begin{titlepage}

\begin{center}

\large{\bf Kendall's Shape Statistics as a Classical Realization of}

\large{\bf Barbour-type Timeless Records Theory approach to Quantum Gravity}

\vspace{.1in}

{\large \bf Edward Anderson}\foo{ea212@cam.ac.uk}

\vspace{.1in}

\large {\em DAMTP Cambridge} \normalsize

\end{center}

\begin{abstract}

I previously showed that Kendall's work on shape geometry is in fact also the geometrical description of 
                         Barbour's relational mechanics' reduced configuration spaces (alias shape spaces). 
I now describe the extent to which        Kendall's subsequent statistical application to e.g. the `standing stones problem' 
realizes further ideas along the lines of Barbour-type timeless records theories, albeit just at the classical level.

\end{abstract}

\vspace{3in} 

\begin{center}

\ni Based on an invited seminar at `Foundations of Physics, Munich 2013'.

\ni Highlights: * Kendall's shape geometry is Barbour's mechanics' reduced configuration space.

\ni             * Kendall's shape statistics is a classical timeless records theory

\ni             * Records theory is a strategy for the Problem of Time

\end{center}

\end{titlepage}

%===========================================================================================================================================
%===========================================================================================================================================
\section{Introduction}\label{Intro}
%===========================================================================================================================================
%===========================================================================================================================================

Julian Barbour proposed a scaled {\it relational particle mechanics (RPM)} in 1982 with Bruno Bertotti BB82 \cite{BB82} and a pure-shape RPM in 2003 \cite{B03}. 
These implement both Temporal Relationalism and Spatial Relationalism \cite{FileR, GrybTh, APoT3} 
in senses that are in accord with Leibniz and Mach's critiques of Newtonian mechanics as explained in Sec 2.  
Moreover, Spatial Relationalism was implemented indirectly in Barbour's formulations of these RPM's, i.e. at the level of unreduced actions.
A natural further question then is what is the specific geometry of these theories' reduced configuration spaces?  

\mbox{ }

\ni I began to study this question from first principles in \cite{06IITricl}, but subsequently found that David Kendall's theory of {\it shape geometry} (Sec 3) 
\cite{Kendall77, Kendall80, Kendall84, Kendall} already covered this in far greater generality in the pure-shape case.
Kendall's work was in fact done in an entirely different context from mechanics: the {\sl statistical} theory of shape. 
Nevertheless, his underlying notion of shape coincides with \cite{FORD, FileR} that used in Barbour's pure-shape RPM.  
Additionally, the cone (in the sense explained in Sec 3) over Kendall's pure-shape geometry turns out to be the reduced configuration space for the scaled RPM \cite{Cones, FileR}.  
Overall, the procedure for constructing a mechanics from a given geometry \cite{Lanczos} to Kendall's geometry (or the cone thereover) 
coincides with the reduction of Barbour's mechanics theories, as outlined in Sec 4.  
Sec 5 then supplies further motivation for Barbour's RPM's from their analogies with GR \cite{B94I, RWR, Kieferbook, FileR}.
These include RPMs manifesting a number of GR's background independent aspects \cite{BI, APoT3, ASoS-AMech} 
and the consequent facets of the Problem of Time in Quantum Gravity \cite{Kuchar92APOTAPOT2}.

\mbox{ } 

\ni Sec 6 then considers one approach to resolving the Problem of Time which Barbour has also worked on \cite{EOT}: the hitherto rather speculative Timeless Records Theory 
(see also \cite{PW83, JZ, GMHH99H03, Page1, Records}).
I then explain (Secs 7 to 10) a new theoretical observation: that Kendall's own further application of shape geometry -- Shape Statistics \cite{Kendall80, Kendall84, 
Kendall85KL87Le87K89, Small, Kendall} -- can furthermore be applied to the RPM case of Records Theory itself.
This is a promising indication of how to render Records Theory a quantitative subject more generally.
Kendall for instance used Shape Statistics to determine whether 
the locations of the standing stones at Land's End in Cornwall contained more alignments than could be put down to to random chance. 
He approached this by the method of sampling in threes, with the probability distribution functions used being based on the geometry of the shape space for the three particles. 
He also applied such Shape Statistics in a more physical setting, involving assessing the supposed evidence for quasars being aligned.

%======================================================================================================================================================
%======================================================================================================================================================
\section{Barbour's Mechanics in Indirect Form}
%======================================================================================================================================================
%======================================================================================================================================================

\ni {\bf Temporal Relationalism} is that that there is no meaningful time for the universe as a whole \cite{B94I, FileR, APoT3} 
as a desirable tenet of background-independence and of closed universes.
This admits the following mathematically sharp implementation. 

\mbox{ }

\ni i)  The action is to neither include any extraneous times -- such as Newtonian time -- nor any extraneous time-like variables (such as the lapse \cite{ADM} in the case of GR).

\ni ii) Time is not to be smuggled into the action in the guise of a label either.
This is attained by postulating {\sl geometrical Jacobi type actions} \cite{Lanczos} 
\beq
S = \sqrt{2}\int\md s\sqrt{W(Q)} \mbox{ } ;
\label{SJ}
\eeq
these so happen to be manifestly parametrization-irrelevant.  
Here 
\beq
\md s := ||\md Q||_{\mbox{\scriptsize \boldmath $M$}}
\label{2}
\eeq 
is the kinetic arc element -- the geometry on the configuration space $\fQ$ of the possible values of the configurations $Q^A$. 
This has a metric ${\mbox{\boldmath $M$}}$ with components ${\mbox{\boldmath $M$}}_{AB}$, which is Riemannian (positive-definite) for Mechanics.
Also the potential factor $W$ for Mechanics is $E - V(Q)$, for potential $V$ and total energy $E$.

A distinguished time that simplifies both \cite{B94I} the equation of motion and the change--momentum relation 
(relational analogue of velocity-momentum relation \cite{ARel2}) then emerges.  
Its form is
\beq
t^{\se\sm} = \int \md s/\sqrt{2W(Q)} \mbox{ } .   
\eeq 
Via (\ref{2}) and $\mbox{d} Q^A$ being a notion of change, this clearly implements Mach's `time is to be abstracted from change' resolution of primary-level timelessness. 
Moreover, this implementation is such that all changes are given an opportunity to contribute \cite{ARel2}.  

\mbox{ } 

\ni 2) {\bf Configurational Relationalism} \cite{BB82, FileR, GrybTh, APoT3}
is that one can take into account the physical irrelevance of a continuous group of transformations $\fG$ acting on the system's configuration space $\fQ$.   
For Mechanics, these are usually translations and rotations of space, though in general Configurational Relationalism also covers physically irrelevant internal transformations, 
as in the most common types of Gauge Theory.  
An indirect implementation of Configurational Relationalism is {\it Best Matching}: 
bringing two configurations into minimum incongruence with each other by application of $\fG$'s group action.  
Then taking into account the linear constraints ensuing from variation with respect to the auxiliary $\fG$-variables sends one to the relationally-desired quotient space $\fQ/\fG$. 
One is to use a cyclic\footnote{This is 
%%%%%%%%%%%%%%%%%%%%%%%%%%%%%%%%%%%%%%%%%%%%%%%%%%%%%%%%%%%%%%%%%%%%%%%%%%%%%%%%%%%%%%%%%%%%%%%%%%%%%%%%%%%%%%%%%%%%%%%%%%%%%%%%%%%%%%%%%%%%%%%%%%%%%%%%%%%%%%%%%%%%%%%%%%%%%%%%%%%%%%%%%%%
`cyclic' in the same Principles of Dynamics sense as `cyclic coordinates' \cite{Lanczos}.}
%%%%%%%%%%%%%%%%%%%%%%%%%%%%%%%%%%%%%%%%%%%%%%%%%%%%%%%%%%%%%%%%%%%%%%%%%%%%%%%%%%%%%%%%%%%%%%%%%%%%%%%%%%%%%%%%%%%%%%%%%%%%%%%%%%%%%%%%%%%%%%%%%%%%%%%%%%%%%%%%%%%%%%%%%%%%%%%%%%%%%%%%%%%
differential presentation $\md g$ for the auxiliary $\fG$-variables \cite{FileR} if this implementation of Configurational Relationalism is to be 
compatible with the above implementation of Temporal Relationalism. 
(The original Lagrange multiplier formulation of the auxiliary $\fG$-variables \cite{BB82, B03} itself breaks the manifest reparametrization invariance.)

For scaled RPM, $\fG$ is the {\it Euclidean group} Eucl($d$) = Tr($d$) $\rtimes$ Rot($d$) 
for Tr the translations, Rot the rotations and $\rtimes$ denoting the semidirect product operation that is well-known from Group Theory.  
The scaled RPM action is\footnote{Here, $I$ runs over the particle labels and $p_I$ are momenta conjugate to $q^I$. 
%%%%%%%%%%%%%%%%%%%%%%%%%%%%%%%%%%%%%%%%%%%%%%%%%%%%%%%%%%%%%%%%%%%%%%%%%%%%%%%%%%%%%%%%%%%%%%%%%%%%%%%%%%%%%%%%%%%%%%%%%%%%%%%%%%%%%%%%%%%%%%%%%%%%%%%%%%%%%%%%%%%%%%%%%%%%%%%%%%%%%%%%%%%
%
The bold quantites are spatial vectors, and I use the calligraphic font to pick out constraints in this Article.} 
%%%%%%%%%%%%%%%%%%%%%%%%%%%%%%%%%%%%%%%%%%%%%%%%%%%%%%%%%%%%%%%%%%%%%%%%%%%%%%%%%%%%%%%%%%%%%%%%%%%%%%%%%%%%%%%%%%%%%%%%%%%%%%%%%%%%%%%%%%%%%%%%%%%%%%%%%%%%%%%%%%%%%%%%%%%%%%%%%%%%%%%%%%%
\be
S_{\mbox{\scriptsize scale}} = \sqrt{2}\int \md s\sqrt{E - V} 
\mbox{ } , \mbox{ }  
\md s^2 := m_I\delta_{IJ} d_{\sbA, \sbB}\bq^Id_{\sbA, \sbB}\bq^J 
\mbox{ }, \mbox{ } 
d_{\sbA, \sbB}\bq^I := \md \bq^I - \md \bA - \md \bB \cr \bq^I \mbox{ } ,
\label{S-scale}
\ee
for $V = V(||\bq^K - \bq^L|| \mbox{ alone})$.

The RPM energy constraint
\beq
{\cal E} := \sumIN\delta_{IJ}\bp^I\bp^J/2m_I + V = E
\label{E}
\eeq
then follows from this action as a primary constraint due to the manifest reparametrization invariance that implements Temporal Relationalism.
Also, the 
\be
\mbox{(zero total momentum of the universe)} \mbox{ } , \mbox{ } \mbox{ } \mbox{\boldmath${\cal P}$} :=  \sumIN\bp_I = 0 \mbox{ } ,  
\ee
\be
\mbox{(zero total angular momentum of the universe)} \mbox{ } , \mbox{ } \mbox{ } \mbox{\boldmath${\cal L}$} := \sumIN\bq^I \cr \bp_I = 0 \mbox{ }   
\ee
follow as secondary constraints from varying with respect to $\bA$ and $\bB$: the outcome of implementing Configurational Relationalism.

For pure-shape RPM, $\fG$ additionally includes the dilations, forming the $d$-dimensional similarity group Sim($d$).
The action then is 
\be
S_{\sp\sss} = \sqrt{2}\int \md s\sqrt{V}                              \mbox{ } , \mbox{ } \mbox{ } 
\md s^2:=m_I\delta_{IJ}\md_{\sbA,\sbB,C}\bq^Id_{\sbA, \sbB, C}\bq^J/I \mbox{ } , \mbox{ } \mbox{ }
\md_{\sA,\sB,C}\bq^I:=\md\bq^I-\md\bA-\md\bB\cr\bq^I+\md C\bq^I       \mbox{ }  
\label{S-ps}
\ee
for $V = V(\mbox{ratios of } ||\bq^K - \bq^L|| \mbox{ alone})$ and $I$ the total moment of inertia.  

This gives similar constraints as before and, from variation with respect to $C$,    
\be
\mbox{(zero total dilational momentum of the universe)} \mbox{ } , {\cal D} := \sumIN\bq^I \cdot \bp_I = 0 \mbox{ } .
\ee
One can instead attempt a direct implementation of Configurational Relationalism, `using gauge-invariant quantities', though this is in general not possible in practise.

%======================================================================================================================================================
%======================================================================================================================================================
\section{Kendall's shape geometry}
%======================================================================================================================================================
%======================================================================================================================================================
%
%FFFFFFFFFFFFFFFFFFFFFFFFFFFFFFFFFFFFFFFFFFFFFFFFFFFFFFFFFFFFFFFFFFFFFFFFFFFFFFFFFFFFFFFFFFFFFFFFFFFFFFFFFFFFFFFFFFFFFFFFFFFFFFFFFFFFFFFFFFFFFFFFFFFFFFFFFFFFFFFFFFFFFFFFFFFFFFFFFFFFFFFF
{           \begin{figure}[ht]
\centering
\includegraphics[width=1.0\textwidth]{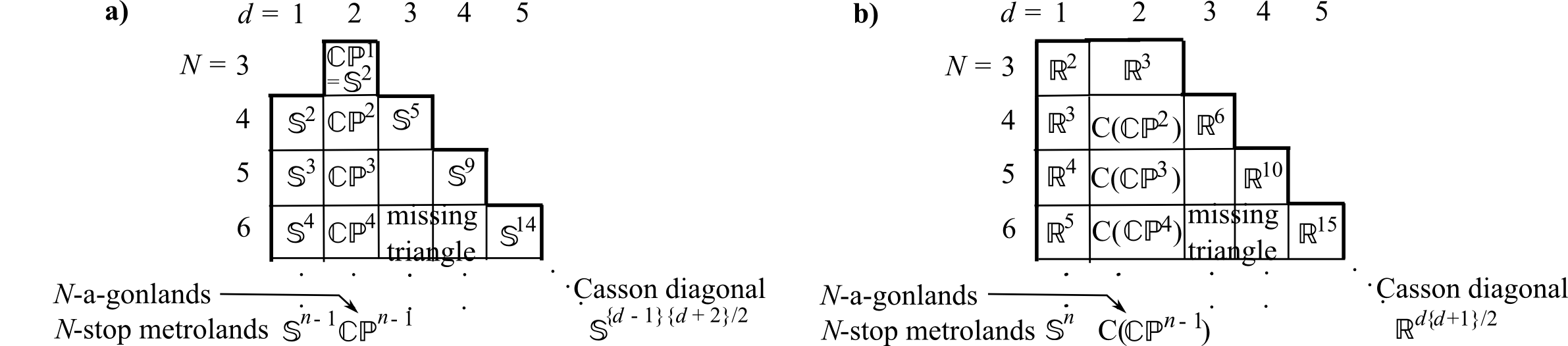}
\caption[Text der im Bilderverzeichnis auftaucht]{  \footnotesize{The distinct a) shape spaces and b) relational spaces at the topological level.  
Only the first 2 columns of each remain straightforward at the metric level, unlike the third topologically simple grouping: the `Casson diagonal' \cite{Kendall} $N = d + 1$.   }   } 
\label{DiscernSS-R} \end{figure}         } 
%FFFFFFFFFFFFFFFFFFFFFFFFFFFFFFFFFFFFFFFFFFFFFFFFFFFFFFFFFFFFFFFFFFFFFFFFFFFFFFFFFFFFFFFFFFFFFFFFFFFFFFFFFFFFFFFFFFFFFFFFFFFFFFFFFFFFFFFFFFFFFFFFFFFFFFFFFFFFFFFFFFFFFFFFFFFFFFFFFFFFFFFF

\ni The set of possible forms of the $N$-point configurations in dimension $d$ -- known as `{\it constellations}' -- form the configuration space $\fQ(N, d) = \mathbb{R}^{Nd}$.  
Kendall then furthermore used the definition that 
\beq
\mbox{(shape space) } \mbox{ } \fS(N, d) := \fQ(N, d)/\mbox{Sim}(d) \mbox{ } ; 
\eeq
note that this coincides with pure-shape RPM's reduced configuration space. 
Moreover, some of these shape spaces have highly tractable mathematics: they are very straightforwardly spheres                   $\mathbb{S}^{N - 2}$ spheres for 1-$d$, 
and, as Kendall demonstrated, \cite{Kendall77, Kendall84, Kendall},                                     complex projective spaces $\mathbb{CP}^{N - 2}$ for 2-$d$.
I term these, respectively, {\it N-stop metrolands} \cite{AF} and {\it N-a-gonlands}. 
The first nontrivial two of the latter are {\it triangleland} \cite{Examples, FileR} and {\it quadrilateralland} \cite{QuadI-II}. 
These results hold at both the topological and metric levels of structure, 
for which the triangleland sphere has the natural spherical metric and the quadrilateralland $\mathbb{CP}^2$ has the natural Fubini--Study metric \cite{Nakahara}.  
Finally, $\mathbb{CP}^1 = \mathbb{S}^2$, rendering triangleland considerably simpler to study than any larger $N$-a-gonland.
Then in this case, one has not only tractable geometry and Methods of Mathematical Physics, but also Kendall's spherical blackboard: Fig \ref{BlackBoard}.a).  

%FFFFFFFFFFFFFFFFFFFFFFFFFFFFFFFFFFFFFFFFFFFFFFFFFFFFFFFFFFFFFFFFFFFFFFFFFFFFFFFFFFFFFFFFFFFFFFFFFFFFFFFFFFFFFFFFFFFFFFFFFFFFFFFFFFFFFFFFFFFFFFFFFFFFFFFFFFFFFFFFFFFFFFFFFFFFFFFFFFFFFFFFFF
{            \begin{figure}[ht]
\centering
\includegraphics[width=1.0\textwidth]{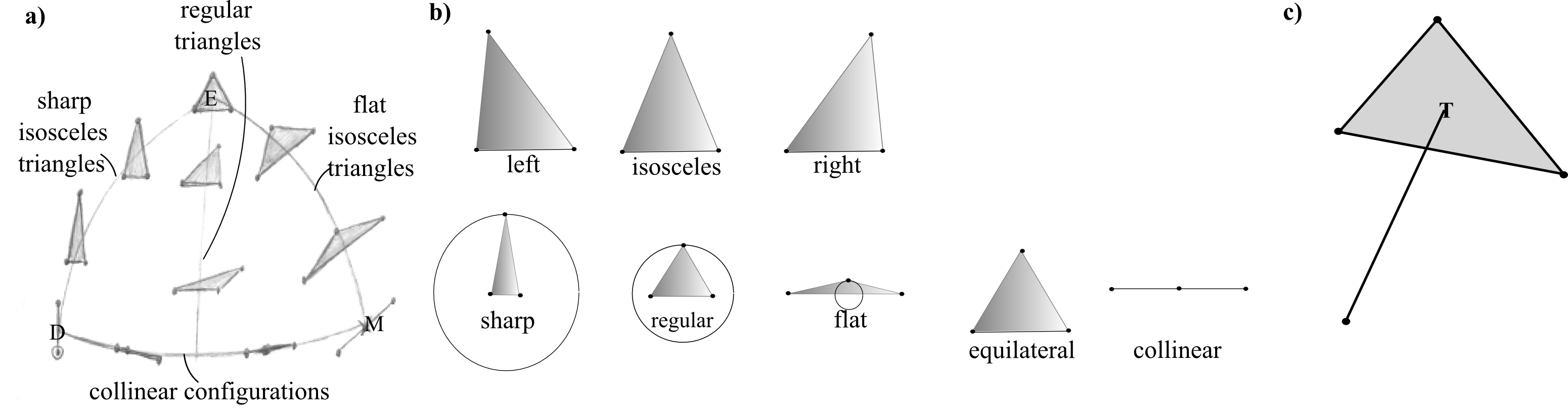}
\caption[Text der im Bilderverzeichnis auftaucht]{        \footnotesize{a) The space of all unlabelled triangles is 1/3 of a hemisphere: Kendall's {\it spherical blackboard}. 
Half of this is drawn here (the other half just contains mirror images).  
b) In dynamics and shape space geometry, a useful description of types of triangles is in terms of  
anisoscelesness: left and right slanting departures from isosceles, 
ellipticity:     sharp and flat departures from regularity, and 
area:            maximal for the equilateral triangle E and minimal for any collinear configuration. 
See Fig 3 for further explanation of this terminology.
Barbour calls sharp configurations `needles', whilst Kendall referred collectively to considerably sharp and flat configurations as `splinters'. 
The double collision D is the sharpest triangle and the merger M is the flattest one.  
c) \cite{QuadI-II} extends a) to quadrilateralland where each point is viewed as an axe (3 + 1 split) as depicted; T is the centre of mass of the triangular `blade' subsystem.}        }
\label{BlackBoard}\end{figure}            }
%FFFFFFFFFFFFFFFFFFFFFFFFFFFFFFFFFFFFFFFFFFFFFFFFFFFFFFFFFFFFFFFFFFFFFFFFFFFFFFFFFFFFFFFFFFFFFFFFFFFFFFFFFFFFFFFFFFFFFFFFFFFFFFFFFFFFFFFFFFFFFFFFFFFFFFFFFFFFFFFFFFFFFFFFFFFFFFFFFFFFFFFFFF

I then defined 
\beq
\mbox{(relational space) } \mbox{ } \fR(N, d) := \fQ(N, d)/\mbox{Eucl}(d) \mbox{ } ;
\eeq
this also coincides with scaled RPM's configuration space.     
The geometry of this is, moreover, the cone over the above shape space \cite{Cones}.\footnote{See \cite{LR97Mont} for some earlier parallel uses of `cone' in 
%%%%%%%%%%%%%%%%%%%%%%%%%%%%%%%%%%%%%%%%%%%%%%%%%%%%%%%%%%%%%%%%%%%%%%%%%%%%%%%%%%%%%%%%%%%%%%%%%%%%%%%%%%%%%%%%%%%%%%%%%%%%%%%%%%%%%%%%%%%%%%%%%%%%%%%%%%%%%%%%%%%%%%%%%%%%%%%%%%%%%%%%%%%
Celestial Mechanics and Molecular Physics.
`Cone' is here used in the following mathematical sense.
A {\it cone} over some topological manifold $\FrMgen$ is denoted by C($\FrMgen$) and takes the form 

\ni \beq
\mbox{C(\FrMgen) = \FrMgen $\times$ [0, $\infty$)/\mbox{ }$\widetilde{\mbox{ }}$} \mbox{ } . 
\eeq
$\widetilde{\mbox{ }}$ \mbox{ } here means that all points of the form \{p $\in$ \FrMgen, 0 $\in [0, \infty)$\} are `squashed' or identified to a single point termed the {\it cone point}, 0. 
Then at the metric level, given a manifold $\FrMgen$ with a metric with line element $\mbox{d} s$, the corresponding cone has a natural metric of form 

\ni\beq
\mbox{d} s^2_{\scc\so\sn\se} := \mbox{d} \rho^2 + \rho^2 \mbox{d} s^2 \mbox{ } . 
\eeq}
%%%%%%%%%%%%%%%%%%%%%%%%%%%%%%%%%%%%%%%%%%%%%%%%%%%%%%%%%%%%%%%%%%%%%%%%%%%%%%%%%%%%%%%%%%%%%%%%%%%%%%%%%%%%%%%%%%%%%%%%%%%%%%%%%%%%%%%%%%%%%%%%%%%%%%%%%%%%%%%%%%%%%%%%%%%%%%%%%%%%%%%%%
% 
This further simplifies for the $N$-stop metroland case via C$(\mathbb{S}^{N - 2}) = \mathbb{R}^{N - 1}$ both topologically and metrically. 
There is no such simplification for $N$-a-gonlands except for triangleland, though that case is not flat (but is conformally flat). 

\mbox{ }

\ni Finally, a consequence of the cone structure is that pure-shape problems occur as subproblems in models with scale. 
Thus considering Kendall's pure-shape case turned out to be key to solving Barbour's scaled RPM also.

%======================================================================================================================================================
%======================================================================================================================================================
\section{Reducing Barbour = building a Mechanics on Kendall}
%======================================================================================================================================================
%======================================================================================================================================================

%FFFFFFFFFFFFFFFFFFFFFFFFFFFFFFFFFFFFFFFFFFFFFFFFFFFFFFFFFFFFFFFFFFFFFFFFFFFFFFFFFFFFFFFFFFFFFFFFFFFFFFFFFFFFFFFFFFFFFFFFFFFFFFFFFFFFFFFFFFFFFFFFFFFFFFFFFFFFFFFFFFFFFFFFFFFFFFFFFFFFFFFF
{            \begin{figure}[ht]
\centering
\includegraphics[width=0.6\textwidth]{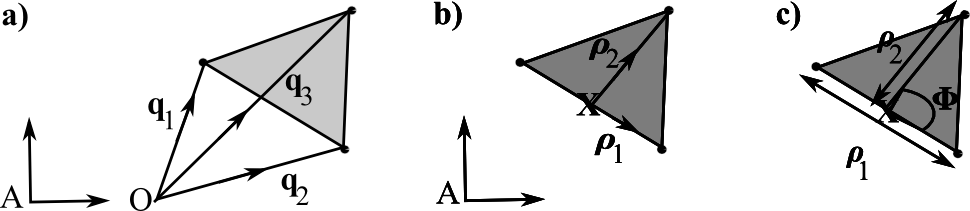}
\caption[Text der im Bilderverzeichnis auftaucht]{        \footnotesize{Progression of coordinate systems for the triangle. 
a) are particle position coordinates relative to an absolute origin O and absolute axes A.  
b) are mass-weighted relative Jacobi interparticle cluster separation coordinates; X denotes the centre of mass of particles 1 and 2.  
N.B. that these coordinates still refer to A.
Then the configuration space radius $\rho := \sqrt{\brho^2_1 + \brho^2_2}$: the square root of the moment of inertia, $I$.    
c) are scaled relational coordinates (ie no longer with respect to any absolute axes either).  
Pure-shape coordinates are then the relative angle $\Phi$ and some function of the ratio $\rho_2/\rho_1$; in particular, $\Theta := 2\,\mbox{arctan}(\rho_2/\rho_1)$.
For normalized mass-weighted relative Jacobi coordinates $\bn_i := \brho_i/\brho$, $aniso = 2 \bn_1\cdot\bn_2$, $area = 2 \{\bn_1\cr\bn_2\}_3$, and $ellip = \bn_2^2 - \bn_1^2$: 
the regular configuration then corresponds to $ellip$ = 0, i.e. to equal partial moments of inertia for the constituent base and `median' subsystems.} }
\label{RPM-coordi} \end{figure}          }
%FFFFFFFFFFFFFFFFFFFFFFFFFFFFFFFFFFFFFFFFFFFFFFFFFFFFFFFFFFFFFFFFFFFFFFFFFFFFFFFFFFFFFFFFFFFFFFFFFFFFFFFFFFFFFFFFFFFFFFFFFFFFFFFFFFFFFFFFFFFFFFFFFFFFFFFFFFFFFFFFFFFFFFFFFFFFFFFFFFFFFFFF

\ni Carry out to completion the relevant set of moves in Fig 3 on the Jacobi arc elements (\ref{S-scale}) and (\ref{S-ps}) in 1- and 2-$d$ for arbitrary $N$ 
to obtain Barbour's RPM in reduced form: the explicit end-product of Best Matching.  
Apply the Jacobi--Synge approach \cite{Lanczos} -- which involves building a natural mechanics from a given metric geometry -- to Kendall's shape spaces and to the cone over these.
I previously showed that these procedures coincide \cite{FORD, Cones, FileR}; 
I term this the {\it Direct = Best-Matched Theorem} after the two types of implementation of Configurational Relationalism.

Thus indeed Kendall's work -- and its straightforward extension by coning -- amounts to having already derived the detailed topological and metric structure of 
the reduced configuration spaces for Barbour's RPM's.
This greatly strengthened the RPM program \cite{Examples, QuadI-II, FileR}.

Note that this coincidence of procedures was by no means guaranteed.  
For instance, Bookstein \cite{Bookstein} gave a distinct metric geometry on the space of shapes.  
I did not study mechanics corresponding to this geometry because it gives material significance to the plane figures themselves, in the form of a `resistance to crushing' criterion. 
Contrast with how relational mechanics considers the constellation of points itself to be the primary entity.  
Also Barbour's Best Matching is a procedure that bears much conceptual similarity to the Procrustes procedure of Shape Geometry \cite{Kendall}; however these each involve 
extremizing a different object.

%=========================================================================================================================================================================================
%=========================================================================================================================================================================================
\section{Further motivation for RPM}
%=========================================================================================================================================================================================
%=========================================================================================================================================================================================

\ni Motivation 1) RPM's are analogous to various significant formulations of GR.

\mbox{ }

\ni 1.i) Wheeler's {\it geometrodynamics} is GR as a dynamical system in which the evolving entities are spatial 3-geometries.  
This is governed by the \cite{ADM}\footnote{The spatial 3-metric $\mh_{ij}$ has determinant $\mh$, 
%%%%%%%%%%%%%%%%%%%%%%%%%%%%%%%%%%%%%%%%%%%%%%%%%%%%%%%%%%%%%%%%%%%%%%%%%%%%%%%%%%%%%%%%%%%%%%%%%%%%%%%%%%%%%%%%%%%%%%%%%%%%%%%%%%%%%%%%%%%%%%%%%%%%%%%%%%%%%%%%%%%%%%%%%%%%%%%%%%%%%%%%%%%
inverse $\mh^{ij}$, covariant derivative $\mD_i$, Ricci scalar $\mR$ and conjugate momentum $\mp^{ij}$.  
$\mN_{abcd}$ is the DeWitt supermetric $\{\mh_{ac}\mh_{bd} - \mh_{ab}\mh_{cd}/2\}/\sqrt{\mh}$.
$\Lambda$ is the cosmological constant.}
%%%%%%%%%%%%%%%%%%%%%%%%%%%%%%%%%%%%%%%%%%%%%%%%%%%%%%%%%%%%%%%%%%%%%%%%%%%%%%%%%%%%%%%%%%%%%%%%%%%%%%%%%%%%%%%%%%%%%%%%%%%%%%%%%%%%%%%%%%%%%%%%%%%%%%%%%%%%%%%%%%%%%%%%%%%%%%%%%%%%%%%%%%%

\beq
\mbox{\it GR Hamiltonian constraint } \mbox{ }     {\cal H} := \mN_{abcd}\mp^{ab}\mp^{cd} - \sqrt{\mh}\{  \mR - 2\Lambda\} = 0
\label{Hamm}
\eeq
\beq
\mbox{ and {\it GR momentum constraint} } \mbox{ } {\cal M}_{i} := - 2\mD_{j}{\mp^{j}}_{i} = 0  \mbox{ } .  
\label{Momm}
\eeq
These are analogous to scaled RPM's ${\cal E}$ and $\mbox{\boldmath${\cal L}$}$ respectively ($\mbox{\boldmath${\cal P}$}$ is rather trivial to remove).  

\mbox{ }

\ni 1.ii) {\it Conformogeometrodynamics} is geometrodynamics' further reformulation in terms of conformal mathematics. 
This has the virtue of decoupling ${\cal M}_i$ from ${\cal H}$, and is then the centrepiece of the initial value problem formulation of GR \cite{Bart-Is}.  
This formulation requires a maximal or CMC slice, given by $\mp = 0$ and $\mp/\sqrt{\mh} = const$ respectively.   
The first of these is furthermore analogous to pure-shape RPM's ${\cal D}$.  

\mbox{ } 

\ni Motivation 2) Moreover, RPM's model a number of aspects of Background Independence \cite{BI, APoT3, ASoS-AMech} and of the subsequent Problem of Time in Quantum Gravity \cite{Kuchar92APOTAPOT2} 
and in Quantum Cosmology in particular \cite{POT-Research}, including exhibiting various further closed-universe effects \cite{FileR}.

\mbox{ } 

\ni 2.i) Temporal and Configurational Relationalism as described in Sec 2 are two aspects of Background Independence that are indeed also manifested by GR.   
The minisuperspace case of GR -- the restriction to homogeneous spatial metrics -- can indeed be cast in terms of an action of form (\ref{SJ}), 
except that now (\ref{2}) is a semi-Riemannian (indefinite) geometry \cite{Magic} and $W = \mR - 2\Lambda$ for GR.
In the case of full GR, the redundant configuration space $\fQ$ = Riem($\Sigma$): the space of Riemannian metrics on a 3-space of fixed topology $\Sigma$.
$\fG$ = Diff($\Sigma$) -- the corresponding spatial 3-diffeomorphisms -- are to be used in the case of geometrodynamics.
Implementing Configurational Relationalism with respect to these by Best Matching gives e.g. the Baierlein--Sharp--Wheeler action \cite{BSW} if the shift auxiliary variable is used, 
or further actions \cite{RWR, AM13} if Temporal Relationalism is to be implemented in tandem with this.
These are still of the general form (\ref{SJ}), except that now integration over space is required in addition to the above interpretations of (\ref{2}) and $W$ carrying over.

Then ${\cal M}_i$ arises from Configurational Relationalism as a minor variant on the usual manner \cite{ADM} in which Diff($\Sigma$) leads to ${\cal M}_i$. 
On the other hand, ${\cal H}$ usually arises in GR from variation with respect to the lapse.
This working has to change, since lapse no longer exists in the relational approach at the primary level.
What happens is that relational actions for GR none the less make up for this non-existence, 
by ${\cal H}$ arising from them as a primary constraint instead, in parallel to how ${\cal E}$ arises in RPM's.

If one chooses instead $\fG$ = Conf($\Sigma$) $\rtimes$ Diff($\Sigma$) or VPConf($\Sigma$) $\rtimes$ Diff($\Sigma$) 
-- where Conf($\Sigma$) are the conformal transformations and VPConf($\Sigma$) are the global volume-preserving conformal transformations -- 
conformogeometrodynamical formulations or theories ensue.
Pure-shape RPM's constraint ${\cal D}$ is furthermore then analogous to conformogeometrodynamics' maximal slice condition.  

\mbox{ } 

\ni 2.ii) Furthermore, relational approach additionally leads to new derivations of both 1.i) and 1.ii) when applied to GR \cite{RWR, ABFKO, AM13}, 
and to various alternative theories and formulations of gravitational theory \cite{Alts}.
Some of the workings involved in this have RPM analogues \cite{FileR}.  

\mbox{ } 

\ni 2.iii) The corresponding GR configuration spaces are the following quotient spaces.
Wheeler's 
\beq
\mbox{Superspace($\Sigma$)                                 := Riem($\Sigma$)/Diff($\Sigma$)} \mbox{ } , 
\eeq
\beq 
\mbox{(conformal superspace) } \mbox{ } \mbox{CS}(\Sigma) := \mbox{Riem}(\Sigma)/\mbox{Conf}(\Sigma) \rtimes Diff(\Sigma) \mbox{ } ,
\eeq 
and the adjunction to the latter of a solitary global volume degree of freedom $\mV$ to form $\mbox{\{CS + V\}}(\Sigma)$.
Compare Sec 3 for RPM analogues; moreover these GR configuration spaces are considerably harder to model than their 1- and 2-$d$ counterparts, rendering them useful model arenas.  
In particular, CS($\Sigma$) is GR's own version of shape space: the space of conformal 3-geometries, whereas \{CS + V\}($\Sigma$) and Superspace($\Sigma$) are globally and 
locally scaled counterparts of this respectively.  
Finally note that 2-$d$ suffices to have a very comprehensive analogy with spatially 3-$d$ GR \cite{FileR}. 
3-$d$ shape geometry is much harder \cite{Kendall} and for very different reasons that GR's own difficulties.  
In this way, 3-$d$ RPM is {\sl far less} suitable as a productive model arena for GR \cite{FileR} than 1- and 2-$d$ RPM are.  

\mbox{ } 

\ni 2.iv) The most well-known facet of the Problem of Time is the Frozen Formalism Problem, a common phrasing for which is that the Wheeler--DeWitt equation -- the quantum wave equation 
\beq
\widehat{\cal H}\Psi = 0 
\eeq
for each whole-universe GR model -- is stationary, i.e. frozen alias timeless.  
Through possessing eq (\ref{E}), RPM's clearly also exhibit a frozen quantum wave equation, 
\beq
\widehat{\cal E}\Psi = 0 \mbox{ } .
\eeq  
Moreover, the Problem of Time is multi-faceted \cite{Kuchar92APOTAPOT2, APoT3}, with RPM's exhibiting around 2/3rds of these \cite{FileR}.

%===============================================================================================================================================================
%===============================================================================================================================================================
\section{Records Theory}
%===============================================================================================================================================================
%===============================================================================================================================================================

\ni One strategy then for dealing with this unexpected frozenness (and the Problem of Time more generally) 
is to take it at face value and see how much Physics one can still do.  
Dynamics or history are now to be {\sl apparent notions} to be constructed from the instant \cite{GMHH99H03} . 
This amounts to supplanting `becoming' with `being' at the primary level.
`Being at a time' is a simpler case to supplant with just `being'.
This is by replacing it by correlations within a single instant between the configuration under study and another that constitutes the `hands of the clock'.   
Supplanting `becoming' itself is more involved \cite{Page1} from a practical perspective.  

\mbox{ }

\ni Theoreticians have differed somewhat both in how to render the notion of timeless records more precise, and in how they envisage the semblance of dynamics may come about.    
Thus there are in fact a number of distinct timeless records approaches \cite{PW83, JZ, Page1, EOT, GMHH99H03, Records}.  
The particular version I consider is as follows at the classical level.

\mbox{ } 

\ni {Records Postulate 1}). Records are information-containing subconfigurations of a single instant that are localized both in space and in configuration space.
This is partly so that they are accurately known and partly so that there can be more than one such to compare.  
For these reasons, Barbour's own insistence on whole-universe configurations is dropped.

\mbox{ } 

\ni {Records Postulate 2)}. Records are furthermore required to contain useful information. 
In this way, such as Information Theory and pattern recognition are thus relevant precursors to Records Theory.

\mbox{ } 

\ni {Records Postulate 3) \cite{FileR}. Records can be tied to atemporal propositions which form a logical structure.
However, this is classically straightforward and thus does not further feature in this Article.  

\mbox{ }

\ni The combined study of the structural levels of Records Postulates 1) to 3) I term {\bf Pre-Records Theory}.  
In a nutshell, this concerns what questions can be asked about whether significant patterns are present in subconfigurations of a single instant.
In support of this, I provided notions of distance and of information as required by Records Postulates 1) and 2) in \cite{FileR} for both Mechanics and GR. 
To make further progress, I next point out that `significant patterns' can be made precise in the sense of {\sl statistically significant} patterns. 
Indeed, I argue in the next four Sections that this is what is required to render Records Theory a quantitative subject, 
using Kendall's Shape Statistics as applied to Barbour's RPM as an example.  

\mbox{ } 

\ni Finally, in order for this to additionally be a minimalistic {\bf Timeless Records Theory}, 
one has to furthermore be able to extract a semblance of dynamics or history from the same-instant information and patterning of 2).

%==========================================================================================================================================================================================
%==========================================================================================================================================================================================
\section{Notions of correlation for RPM's}
%==========================================================================================================================================================================================
%==========================================================================================================================================================================================

I first present some very elementary Probability and Statistics notions, 
alongside relational modifications to them which render them suitable for addressing questions about RPM configurations.
For random variables X, Y, 
\beq
\mbox{(Pearson's correlation coefficient) } \mbox{ }  \rho_{\sP} := \mbox{Cov(X,   Y)}\big/ \sqrt{\mbox{Var(X)\,Var(Y)}} \mbox{ } , 
\label{Pearson}
\eeq
i.e. the normalization of the covariance by the square roots of the individual random variables' variances. 

\mbox{ } 

\ni Suppose then that we are given a constellation of $N$ points in 2-$d$, the physical content of which we term an $N$-a-gon in this Article, and which is a 2-$d$ RPM configuration.
Is it relational to assess this for collinearity using $\rho_{\sP}$? 
Preliminarily note that the particular form of $\rho_{\sP}$ that I choose to exhibit is the one which is invariant under exchange of dependent and independent variable statuses of X and Y.  
This is relevant since such a difference is immaterial in the RPM application.  
However, there is another reason by which the answer is no.
For, as is very well known, the upward- and downward-pointing lines correspond to {\sl opposite} extremes of $\rho_{\sP}$.
But these configurations are relationally indiscernible, so $\rho_{\sP}$ cannot be entirely relational.\footnote{The everyday {\it rank correlation tests}' statistics 
%%%%%%%%%%%%%%%%%%%%%%%%%%%%%%%%%%%%%%%%%%%%%%%%%%%%%%%%%%%%%%%%%%%%%%%%%%%%%%%%%%%%%%%%%%%%%%%%%%%%%%%%%%%%%%%%%%%%%%%%%%%%%%%%%%%%%%%%%%%%%%%%%%%%%%%%%%%%%%%%%%%%%%%%%%%%%%%%%%%%%%%%%%%
also immediately fail to be rotationally relational. 
This is since any two data points can be rotated into a tie, and the procedures for these tests exclude ties.} 
%%%%%%%%%%%%%%%%%%%%%%%%%%%%%%%%%%%%%%%%%%%%%%%%%%%%%%%%%%%%%%%%%%%%%%%%%%%%%%%%%%%%%%%%%%%%%%%%%%%%%%%%%%%%%%%%%%%%%%%%%%%%%%%%%%%%%%%%%%%%%%%%%%%%%%%%%%%%%%%%%%%%%%%%%%%%%%%%%%%%%%%%%%%  
%
In fact, variance and covariance are translation-invariant and the normalization used in $\rho_{\sP}$ renders this scale-invariant as well, but it indeed fails to be rotationally invariant.

How can one remedy this for use in relational problems like Kendall's standing stones problem or the study of snapshots from the $N$-a-gonland mechanics? 
One such approach follows from the 
\beq
\mbox{(covariance matrix) } \mbox{ }  \bV := \mbox{\Huge(}\stackrel{ \mbox{ } \mbox{ } \mbox{Var(X)}  \mbox{ } \mbox{ } \mbox{ }  \mbox{Cov(X, Y)}}
                            {\mbox{Cov(X,Y)}                    \mbox{ }  \mbox{ }                                                  \mbox{Var(Y)}   }\mbox{\Huge)}
\eeq
being a 2-tensor under the corresponding Rot(2) = $SO(2)$ rotations.  
Thus consider the corresponding invariants: det$\,\bV$ = Var(X)Var(Y) -- Cov(X,Y)$^2$ and tr$\,\bV$ = Var(X) + Var(Y).
In particular, $\sqrt{\mbox{det}\,\bV}$/tr$\,\bV$ is a scale-invariant ratio of translation-and-rotation invariant quantities 
and is therefore a relationally invariant base-object for single-line linear correlations.
This can be repackaged as
\beq
\mbox{(relational correlation coefficient) } \mbox{ }{\rho}_{\sR\se\sll} := 2{\sqrt{1 - \rho^2_{\sP}}}\big/\{\omega + \omega^{-1}\}
\label{RelCorr}
\eeq
for $\omega := \sqrt{\mbox{Var(X)}/\mbox{Var(Y)}}$.  
[Also note that taking functions of this may allow for more flexibility as regards passing further hurdles from Statistics.] 

\mbox{ } 

\ni Moreover, (\ref{RelCorr}) must be a function of the basis of shape quantities for triangleland. 
If the data is taken to represent the Jacobi vectors, $\sqrt{\mbox{det}\,\bV} \propto I \,\, area$, as one might anticipate from how $area$ is attained by noncollinearity.  
Additionally, tr$\,\bV \propto I\{1 - aniso\}$.  
Thus, all in all, 
\beq
{\rho}_{\sR\se\sll} \propto {area}/\{1 - aniso\} \mbox{ } .  
\eeq
This forewarns us that normalizing by use of $I$ itself (c.f. Fig 3) is not natural in {\sl statistical} investigations, unlike in the preceding Secs' mechanical calculations.

%=====================================================================================================================================================================================
%=====================================================================================================================================================================================
\section{Shape Statistics of clumping}\label{Clump}
%=====================================================================================================================================================================================
%=====================================================================================================================================================================================

The previous Sec just considers an overall pattern: whether the entirety of the points in the constellation fit a single straight line well.

Let us next consider the further also fairly well-known case of clumping.  
Astrophysical examples of clumping include tight binary stars, globular clusters, galaxies, and voids: {\sl absense} of clumping}.   
This is `ratios of relative separations' information, which can already be modelled by particles in 1-$d$.  
This has the further advantages of involving more detailed information which can be attributed locally and to subsystems.  
In this manner, clumping statistics can be used to assess the significance of more detailed aspects of patterns, 
as well as applying to localized subconfigurations in the context of Records Theory.  
E.g. Roach \cite{Roach} provided a discrete statistical study of clumping; this can in turn be interpreted in terms of coarse-grainings of RPM configurations.

Also note that the previous Sec's fitting a line concerns relative angle information between the points in the constellation, in distinction to ratios of relative separations information.
These are two distinct types of shape information.
Moreover, in 1-$d$, only the latter type of shape information exists.
Thus 1-$d$ RPM's can be used to study ratios of relative separations information {\sl in isolation from} \cite{AF} the more complicated assessment of relative angle information. 
On the other hand, 2-$d$ RPM's suffice to exhibit both, in fact in a 1 : 1 proportion.
This follows most readily from the $\mathbb{C}$ presentation underlying the $\mathbb{C}^{N - 2}$ having one independent ratio information magnitude per independent relative angle phase.  
Moreover, the previous Sec's approach is but one particular case of assessing relative angle information, and one which uses the whole data set rather than localized subconfigurations.
A more general and detailed method for assessing relative angle information, which furthermore focuses on subconfigurations, is laid out in the next Section.  

%FFFFFFFFFFFFFFFFFFFFFFFFFFFFFFFFFFFFFFFFFFFFFFFFFFFFFFFFFFFFFFFFFFFFFFFFFFFFFFFFFFFFFFFFFFFFFFFFFFFFFFFFFFFFFFFFFFFFFFFFFFFFFFFFFFFFFFFFFFFFFFFFFFFFFFFFFFFFFFFFFFFFFFFFFFFFFFFFFFFFFFFF
{            \begin{figure}[ht]
\centering
\includegraphics[width=0.7\textwidth]{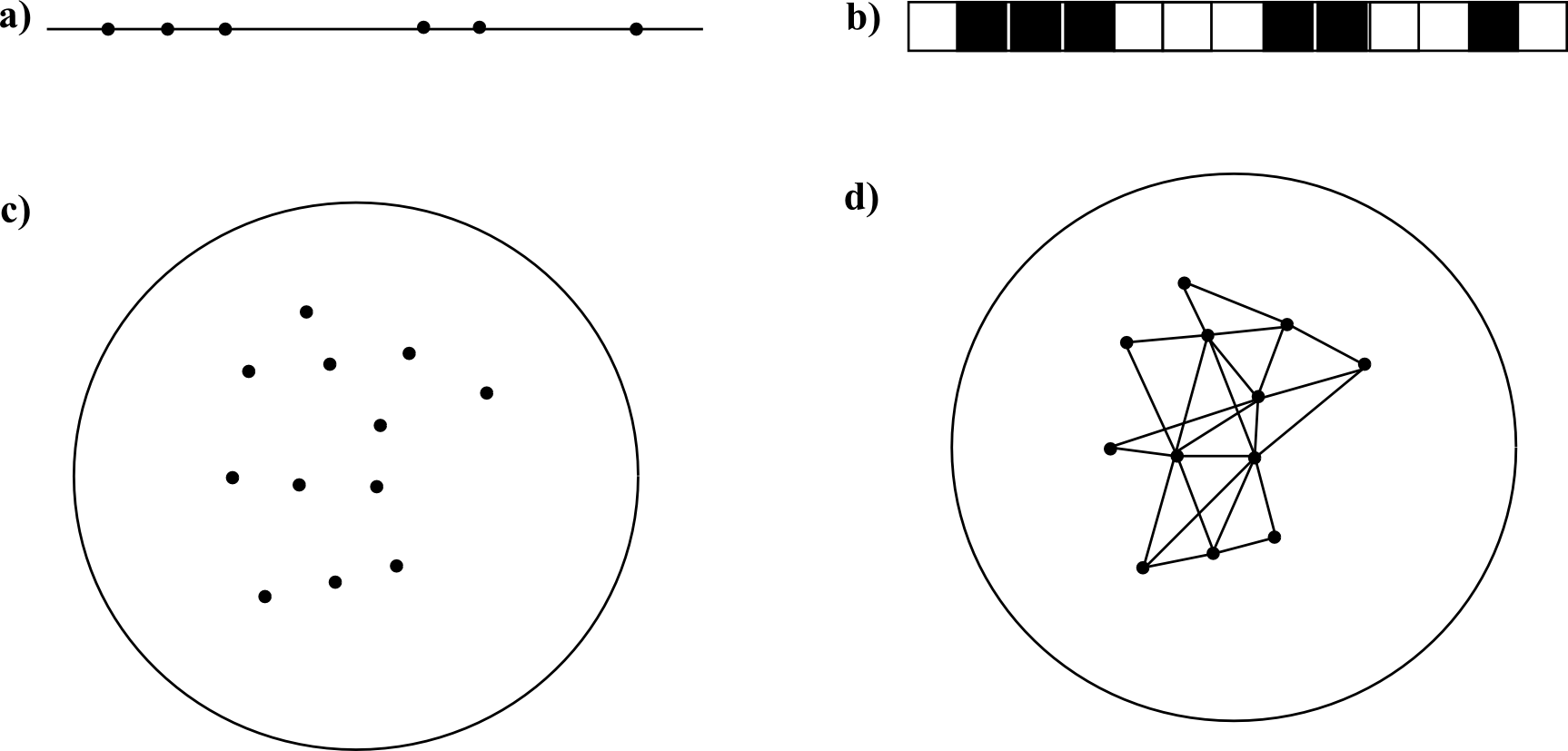}
\caption[Text der im Bilderverzeichnis auftaucht]{        \footnotesize{a) Clumping in 1-$d$ and b) a discrete model of it as per \cite{Roach}.  
c) Shape data in 2-$d$.  This could consist of e.g. standing stones or of the particle positions in an RPM.  
d) Is the number of almost-collinear triangles present accountable for by coincidence or is it statistically significant pattern? }    }
\label{Clumping-Kendall} \end{figure}          }
%FFFFFFFFFFFFFFFFFFFFFFFFFFFFFFFFFFFFFFFFFFFFFFFFFFFFFFFFFFFFFFFFFFFFFFFFFFFFFFFFFFFFFFFFFFFFFFFFFFFFFFFFFFFFFFFFFFFFFFFFFFFFFFFFFFFFFFFFFFFFFFFFFFFFFFFFFFFFFFFFFFFFFFFFFFFFFFFFFFFFFFFF

%=====================================================================================================================================================================================
%=====================================================================================================================================================================================
\section{Kendall's Shape Statistics}\label{DrNo}
%=====================================================================================================================================================================================
%=====================================================================================================================================================================================

Indeed, seeking a single regression line for the best fit of a whole data set as in Sec 7 is not always appropriate. 
This is made clear by the study of the distribution of standing stones or quasars.
Here, one is to consider more sophisticated geometrical propositions about $N$-a-gon constellations. 
E.g. assessing whether there is a disproportionate number of approximately-collinear triples of points -- 
a more detailed relative angle information criterion and indeed tied to subconfigurations.  
This is addressed by sampling in threes.
The standing stones problem was first posed as a problem of this kind to be addressed by Geometrical Statistics by Broadbent \cite{Broadbent}.  

\mbox{ } 

\ni Some relevant details of the statistical tests used are as follows. 
One of the approaches by Kendall and collaborators \cite{Kendall80, Kendall84, Kendall85KL87Le87K89, Small, Kendall} 
involves the assumption that the standing stones lie within a compact convex polygon (`the Cornish coastline' for the particular Land's End standing stones problem).  
Detail of the compact convex polygon then enters the test's outcome.  
However, this restriction is made in order to to cope with uniform independent identically-distributed distributions (much as quantum physicists often perform `normalization by boxing'). 
Moreover, other approaches by Kendall and collaborators use distributions that tail off; in these cases, involvement of a `coastline' ceases to be necessary.

The probability distributions in question then live on the shape space of the triangles that one is sampling with.  
As per Sec 3, this is (a piece of) $\mathbb{S}^2$ for which geometry and subsequent mathematical methods are well-understood.
In this manner, one is dealing with a {\it Geometrical Statistics}.
A fortiori, it is a {\it Shape Statistics} in accord with the following distinction.  
Geometrical statistics concerns applying Probability and Statistics techniques to sample spaces that are differentiable manifolds \cite{Small, Kendall}, 
involving suitable notions of $\sigma$-field, geometrical measure, change of variables formulae and taking isometries into account. 
Shape Statistics, however, concerns a differentiable (in general stratified)\foo{A {\it stratified manifold} is a structure that is more general than a manifold, 
%%%%%%%%%%%%%%%%%%%%%%%%%%%%%%%%%%%%%%%%%%%%%%%%%%%%%%%%%%%%%%%%%%%%%%%%%%%%%%%%%%%%%%%%%%%%%%%%%%%%%%%%%%%%%%%%%%%%%%%%%%%%%%%%%%%%%%%%%%%%%%%%%%%%%%%%%%%%%%%%%%%%%%%%%%%%%%%%%%%%%%%%%%%
within which dimension can vary from point to point but which still fits together according to topologically `nice' rules \cite{AConfig}.}
%%%%%%%%%%%%%%%%%%%%%%%%%%%%%%%%%%%%%%%%%%%%%%%%%%%%%%%%%%%%%%%%%%%%%%%%%%%%%%%%%%%%%%%%%%%%%%%%%%%%%%%%%%%%%%%%%%%%%%%%%%%%%%%%%%%%%%%%%%%%%%%%%%%%%%%%%%%%%%%%%%%%%%%%%%%%%%%%%%%%%%%%%%%
manifold the points of which can, a furthermore, be interpreted as `shapes in space'.

Finally, a suitable conceptual notion for assessing the approximate collinearity of triples of points is Kendall's notion of $\epsilon$-bluntness (see Fig \ref{Ecoll}.a). 
%
%FFFFFFFFFFFFFFFFFFFFFFFFFFFFFFFFFFFFFFFFFFFFFFFFFFFFFFFFFFFFFFFFFFFFFFFFFFFFFFFFFFFFFFFFFFFFFFFFFFFFFFFFFFFFFFFFFFFFFFFFFFFFFFFFFFFFFFFFFFFFFFFFFFFFFFFFFFFFFFFFFFFFFFFFFFFFFFFFFFFFFFFFF
{            \begin{figure}[ht]
\centering
\includegraphics[width=0.9\textwidth]{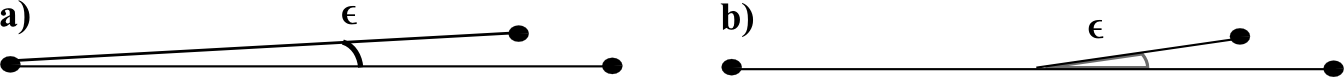}
\caption[Text der im Bilderverzeichnis auftaucht]{        \footnotesize{Meanings in 2-$d$ space for 
a) Kendall's \cite{Kendall} $\epsilon$-blunt notion of collinearity: angle at any vertex $\leq \epsilon$ is significant.
b) My notion of $\epsilon$-collinear that is adapted to the dynamically useful Jacobi variables: angle $\Phi \leq \epsilon$ is significant.  }    }  
\label{Ecoll}\end{figure}            }
%FFFFFFFFFFFFFFFFFFFFFFFFFFFFFFFFFFFFFFFFFFFFFFFFFFFFFFFFFFFFFFFFFFFFFFFFFFFFFFFFFFFFFFFFFFFFFFFFFFFFFFFFFFFFFFFFFFFFFFFFFFFFFFFFFFFFFFFFFFFFFFFFFFFFFFFFFFFFFFFFFFFFFFFFFFFFFFFFFFFFFFFFF

\mbox{ } 

\ni Kendall additionally generalized such methods from sampling in triples to sampling in $M$-tuples in 2-$d$ \cite{Kendall84}.

%=========================================================================================================================================================================================
%=========================================================================================================================================================================================
\section{Application to RPM Records Theory}
%=========================================================================================================================================================================================
%=========================================================================================================================================================================================

{\bf Clumping--Kendall paradigm for RPM Records Theory}.  
Roach's clumping and Kendall's alignment in threes cover the detection of {\sl discernible patterns} in the ratios of relative separations and relative angle types of shape information in 
the $N$-a-gon configurations of RPM.  
Furthermore, it is discernible patterns that distinguish certain subconfigurations of an instant as a pre-record. 
Thus these statistical techniques constitute quantitative means of determining whether a given configuration is a meaningful pre-record rather than a purely random subconfiguration.
Between Shape Geometry furnishing the notion of distance needed for Records Postulate 1) and Shape Statistics furnishing the notions of useful information and correlation needed for 
Records Postulate 2) as a corollary of the `Direct = Best-Matched' Theorem,  we have a working theory of classical pre-records.

\mbox{ }

\ni This pioneers the use of a wider variety of notions of pattern assessment so as to render {\sl any} physical theory's Records Theory formulation quantitatively meaningful.
See \cite{ASoS-AMech} for outlines of examples of notions of shape, Shape Mechanics, Shape Statistics and Records Theories in a wider range of settings.

\mbox{ }

\ni Let us comment in more detail on Kendall's approach from an RPM perspective.
N.B. that the method of sampling in threes is based on probing with the 2-$d$ model's minimal-sized nontrivially-relational subconfigurations: the constituent triangles.
The subset of methods involving fitting within a convex polygon are not per se a restriction on RPM instant configurations, since one can always rescale these to fit within.
However, detail of the convex shape in question entering the conclusion would now constitute an unwanted vestige of absolute structure.  
Thus the other subset of methods involving probability distributions that tail off instead is particularly welcome in the RPM context, 
as a freeing from such a background-independent imprint.

The pre-records found peak about the  real projective space $\mathbb{RP}^{N - 2}$ `equator' of collinearity 
that each $N$-a-gonland shape space $\mathbb{CP}^{N - 2}$ possesses \cite{FileR}.  
Since RPM's hitherto studied in detail have involved small particle numbers \cite{Examples, QuadI-II, FileR}, 
I point out that the 3 points of a triangle are too few for this application to Records Theory. 
This is because sampling in threes means that a single triangle's worth of data plays the role of the statistically-meaningless sample size of 1.
Nontriviality in this application thus starts with quadrilateral configurations \cite{QuadI-II}, which allow for sampling with up to 4 constituent triangles.
Moreover, one needs a constellation containing somewhat more points than that in order to attain a {\sl statistically significant} sample size.

\mbox{ } 

\ni Finally, as regards obtaining a semblance of history from records, now significant results for different values of $\epsilon$ carry different implications \cite{Broadbent}. 
Were they laid out skillfully by the epoch's standards for e.g. astronomical or religious reasons ($\epsilon \leq 10$ minutes of arc), 
or were they just the markers of routes or plots of land ($\epsilon \leq 1$ degree)?

%========================================================================================================================================================================================
%========================================================================================================================================================================================
\section{Conclusion}
%========================================================================================================================================================================================
%========================================================================================================================================================================================

Barbour's relational particle mechanics (RPM) are of considerable interest both as theories of mechanics which implement Leibnizian and Machian features 
                                                                            and as toy models of many aspects of GR-as-geometrodynamics.
Direct Jacobi--Synge construction of the mechanics corresponding to Kendall's shape space geometry and to the cone thereover produces the same mechanics as reducing Barbour's 
pure-shape and scaled RPM's respectively, which he formulated indirectly instead: using Best Matching. 
Subsequent application of Shape Statistics to Theoretical Physics is new. 
In the present article, I have shown that this is ready-built for use in the case of the 1- and 2-$d$ RPM's -- the RPM's that are {\sl productive} model arenas of GR-as-geometrodynamics.
This can be used to quantify whether given subconfigurations are records, a matter of interest in timeless (and histories --- see below) approaches to Quantum Gravity.
I leave expressing the Shape Statistics quantifiers themselves in terms of these for a future occasion.

\mbox{ }

\ni Instead, I pose a difficult and interesting question.  
This follows from GR having its own analogue of shape space: CS($\Sigma$), 
and globally and locally scaled counterparts thereof: \{CS + V\}($\Sigma$) and Superspace($\Sigma$) respectively.  
The question then is what are the corresponding notions of (scale and) Shape Statistics, 
which one would use to quantitatively detect records within (conformo)geometrodynamical subconfigurations?  

\mbox{ } 

\ni Some intermediate steps toward this are as follows (see \cite{FileR, AConfig} for details and references).
Minisuperspace (homogeneous GR) and inhomogeneous perturbations about homogeneous GR are rather simpler models than full GR.

\mbox{ } 

\ni 1) Diagonal minisuperspace has a  shape space -- the space of anisotropies -- that is flat, so it is too simple for nontrivial application of Geometrical Statistics \cite{AConfig}.

\ni 2) At the other extreme, full GR's shape space --- CS($\Sigma$) --- is both infinite-$d$ and a stratified manifold,  
though some of Kendall's work \cite{Kendall} makes it clear that Geometrical Statistics can indeed transcend to stratified manifold geometries. 

\ni 3) The inhomogeneous perturbation model is intermediate in complexity between the previous two cases, 
and can furthermore be seen as a replacement of the point particles of RPM's by small inhomogeneous lumps within a GR framework. 
By this, and by CS($\Sigma$) being a shape space, the name `Shape Statistics' indeed continues to be merited for this example by the argument given in Sec 9. 
Furthermore, this line of enquiry for 3) could lead to new methods and insights as regards the analysis of the detailed data recently gathered by the Planck satellite.  

\mbox{ } 

\ni Finally, of course, one's ultimate interest as regards Quantum Gravity and the Records Theory approach to this is, additionally, quantum-mechanical.  
Barbour suggested \cite{EOT} the formation of tracks by $\alpha$-particles in bubble chambers as a paradigm for timeless records formation from which a sense of dynamics or history 
could be extracted.  
However, it may well be far more typical for decoherence to leave most to all information in an irretrievable state, 
as suggested e.g. by the Joos--Zeh paradigm of a dust particle decohering due to the microwave background photons \cite{JZ}. 
Barbour also conjectured that quantum probability density function `mist' might peak in some geometrically distinguished region of configuration space, 
whose configurations happen to be meaningful records \cite{EOT}. 
Unfortunately for this conjecture, the concrete examples of small classical and quantum RPM's that I have worked out so far do not support it \cite{FileR}.  
N.B. that Records Theory is useful not only in purely timeless approaches but also in approaches which assume a sense of history \cite{GMHH99H03, POT-Research}.
This further adds to the value of the present research.  

\mbox{ } 

\ni {\bf Acknowledgements} I thank the Conference Organizers for inviting me to speak at the `Colloquium in Julian Barbour's honour' part of this Conference, 
and I thank Julian Barbour also for many discussions over the years.
I also thank Christopher Small, Huiling Le, the Anonymous Referee and many people at the `Conformal Nature of the Universe' Conference at the Perimeter Institute in 2012 for discussions.  
And Grant FQXi-RFP3-1101 from the Foundational Questions Institute (FQXi) Fund, administered by 
Silicon Valley Community Foundation, Theiss Research and the CNRS, hosted with Marc Lachi$\grave{\me}$ze-Rey at APC.

%=====================================================BIBLIOGRAPHY================================================================================================================


\begin{thebibliography}{99}
%=================================================================================================================================================================================

\footnotesize

\bibitem{BB82}               J.B. Barbour and B. Bertotti, Proc. Roy. Soc. Lond. {\bf A382} 295 (1982). 

\bibitem{B03}                J.B. Barbour, Class. Quant Grav. \textbf{20}, 1543 (2003), gr-qc/0211021.

\bibitem{FileR}              E. Anderson, version 3 of arXiv:1111.1472.  

\bibitem{GrybTh}             S.B. Gryb, (Ph.D. Thesis, Waterloo, Canada 2011), arXiv:1204.0683.  

\bibitem{APoT3}              E. Anderson, arXiv:1409.4117. 

\bibitem{06IITricl}          E. Anderson  Class. Quant Grav. {\bf 23} 2491 (2006), gr-qc/0511069;
%
                             {\bf 24} 5317 (2007), gr-qc/0702083. 

\bibitem{Kendall77}          D.G. Kendall, Adv. Appl. Probab. {\bf 9} 428 (1977).  

\bibitem{Kendall80}          D.G. Kendall and W.S. Kendall,  Adv. Appl. Probab. {\bf 12} 380 (1980).

\bibitem{Kendall84}          D.G. Kendall, Bull. Lond. Math. Soc. {\bf 16} 81 (1984). 

\bibitem{Kendall}            D.G. Kendall, D. Barden, T.K. Carne and H. Le, {\it Shape and Shape Theory} (Wiley, Chichester 1999).  

\bibitem{FORD}               E. Anderson, Class. Quant Grav. {\bf 25} 025003 (2008), arXiv:0706.3934.

\bibitem{Cones}              E. Anderson, arXiv:1001.1112.

\bibitem{Lanczos}            C. Lanczos, {\it The Variational Principles of Mechanics} (University of Toronto Press, Toronto 1949). 

\bibitem{B94I}               J.B. Barbour, Class. Quant. Grav. {\bf 11} 2853 (1994).

\bibitem{RWR}                J.B. Barbour, B.Z. Foster and N. \'{O} Murchadha, Class. Quant Grav. {\bf 19} 3217 (2002), gr-qc/0012089.  

\bibitem{Kieferbook}         C. Kiefer, {\it Quantum Gravity} (Clarendon, Oxford 2004).  

\bibitem{BI}                 E. Anderson, arXiv:1310.1524.  

\bibitem{ASoS-AMech}         E. Anderson, arXiv.1412.0239;
%
                             arXiv:1505.00488. 
						 
\bibitem{Kuchar92APOTAPOT2}  K.V. Kucha\v{r}, in {\it Proceedings of the 4th Canadian Conference on General Relativity and Relativistic Astrophysics} 
                             ed. G. Kunstatter, D. Vincent and J. Williams (World Scientific, Singapore 1992); 
%
                             E. Anderson, in {\it Classical and Quantum Gravity: Theory, Analysis and Applications} ed. V.R. Frignanni (Nova, New York 2011), arXiv:1009.2157; 
%							 
                             Annalen der Physik, {\bf 524} 757 (2012),  arXiv:1206.2403.   

\bibitem{EOT}                J.B. Barbour, Class. Quant. Grav. {\bf 11} 2875 (1994); 
%  
			                 {\it The End of Time} (OUP, Oxford 1999).

\bibitem{PW83}               D.N. Page and W.K. Wootters, Phys. Rev. {\bf D27}, 2885 (1983).

\bibitem{JZ}                 E. Joos and H.D. Zeh, Z. Phys. {\bf B59} 223 (1985).
							 							 
\bibitem{GMHH99H03}	         M. Gell-Mann and J.B. Hartle, Phys. Rev. {\bf D47} 3345 (1993); 
% 
                             J.J. Halliwell, Phys. Rev. {\bf D60} 105031 (1999), quant-ph/9902008.
%
                             in {\it The Future of Theoretical Physics and Cosmology  
                             (Stephen Hawking 60th Birthday Festschrift Volume)} ed. G.W. Gibbons et al (CUP, Cambridge 2003), gr-qc/0208018.  

\bibitem{Page1}              D.N. Page, Int. J. Mod. Phys. {\bf D5} 583 (1996), quant-ph/9507024.

\bibitem{Records}            E. Anderson, Int. J. Mod. Phys. {\bf D18} 635 (2009), arXiv:0709.1892;   
%
                             in {\it Proceedings of the Second Conference on Time and Matter}, ed. M. O'Loughlin et al (UNGP, Nova Gorica, Slovenia 2008), arXiv:0711.3174. 
 							 							 
\bibitem{Kendall85KL87Le87K89} D.G. Kendall, Adv. Appl. Prob. {\bf 17} 308 (1985);
%
                             D.G. Kendall and H. Le,  Adv. Appl. Prob. {\bf 19} 896 (1987); 
%
                             H. Le, Math. Proc. Camb. Phil. Soc. {\bf 102} 587 (1987); 	
%
                             D.G. Kendall, Statistical Science {\bf 4} 87 (1989).

\bibitem{Small}              C.G.S. Small, {\it The Statistical Theory of Shape} (Springer, New York 1996).  

%===================================================== BODY ============================================================================================							 

\bibitem{ADM}                R. Arnowitt, S. Deser and C.W. Misner, ``The Dynamics of General Relativity", 
                             in {\it Gravitation: An Introduction to Current Research} ed. L. Witten (Wiley, New York 1962), arXiv:gr-qc/0405109.  

\bibitem{ARel2}              E. Anderson, arXiv:1209.1266. 
							 
\bibitem{AF}                 E. Anderson and A. Franzen, Class. Quantum Grav. {\bf 27} 045009 (2010), arXiv:0909.2436.  
							 
\bibitem{Examples}           E. Anderson, Gen. Rel. Grav. {\bf 43} 1529 (2011), arXiv:0909.2439;
%
                             Class. Quantum Grav. {\bf 28} 065011 (2011), arXiv:1003.1973; 
%
                             arXiv:1005.2507.  

\bibitem{QuadI-II}           E. Anderson, Int. J. Mod. Phys. {\bf D23} 1450014 (2014), arXiv:1202.4186; 							 
%							 							 
                             E. Anderson and S.A.R. Kneller, Int. J. Mod. Phys. {\bf D23} 1450052 (2014), arXiv:1303.5645.  

\bibitem{Nakahara}            M. Nakahara, {\it Geometry, Topology and Physics} (Institute of Physics Publishing, London 1990).   
							 
\bibitem{LR97Mont}           R. Montgomery, Nonlin. {\bf 9} 1341 (1996); 
%
                             R.G. Littlejohn and M. Reinsch, Rev. Mod. Phys. {\bf 69} 213 (1997).
							 
\bibitem{Bookstein}          F.L. Bookstein, Statistical Science {\bf 4} 99 (1989).
					 
\bibitem{Bart-Is}            R. Bartnik and J. Isenberg, in {\it The Einstein Equations and the Large Scale Behavior of Gravitational Fields} 
                             ed. P. Chrusciel and H. Friedrich (Birkh\"{a}user, Basel 2004).

\bibitem{POT-Research}       E. Anderson, Class. Quant. Grav. {\bf 29} 235015 (2012), arXiv:1204.2868; 
%
                             Class. Quant. Grav {\bf 31} 025006 (2014) arXiv:1305.4685;
%
                             Ann. N.Y. Acad. Sci. {\bf 1326} 42 (2014) arXiv:1306.5816;
%                             
                             arXiv:1403.7583. 
 
\bibitem{Magic}              C.W. Misner, in {\it Magic Without Magic: John Archibald Wheeler} ed. J. Klauder (Freeman, San Francisco 1972).
							 							 
\bibitem{BSW}                R.F. Baierlein, D. Sharp and J.A. Wheeler, Phys. Rev. {\bf 126} 1864 (1962). 

\bibitem{AM13}               E. Anderson and F. Mercati, arXiv:1311.6541. 
 
\bibitem{ABFKO}              E. Anderson, Stud. Hist. Phil. Mod. Phys. {\bf 38} 15 (2007), gr-qc/0511070; 
%	
                             E. Anderson, J.B. Barbour, B.Z. Foster, B. Kelleher and N. \'{O} Murchadha, Class. Quant Grav {\bf 22} 1795 (2005), gr-qc/0407104.
						 							 
\bibitem{Alts}               E. Anderson, Gen. Rel. Grav. {\bf 36} 255 (2004), gr-qc/0205118; 
%
                             E. Anderson, J.B. Barbour, B.Z. Foster and N. \'{O} Murchadha, Class. Quant Grav. {\bf 20} 157 (2003), gr-qc/0211022; 
%
							 T. Budd and T. Koslowski, arXiv:1107.1287;  
%
                             H. de A. Gomes and T. Koslowski, Class. Quant. Grav. {\bf 29} 075009 (2012), arXiv:1101.5974;  
%
                             arXiv:1206.4823;
%
                             S.B. Gryb and F. Mercati, Phys. Rev. {\bf D87} 064006 (2013) arXiv:1209.4858.  
							 
\bibitem{Roach}              S.A. Roach, {\it The Theory of Random Clumping} (Methuen, London 1968).  
						 							 																		 
\bibitem{Broadbent}          S. Broadbent, J.R. Statist. Soc. {\bf A 143} 109 (1980).  

\bibitem{AConfig}            E. Anderson, arXiv:1503.01507. 
							 
\end{thebibliography}
\end{document}